\documentclass{appolb}
\usepackage{epsfig}

\begin{document}

\title{Tsallis Fits to $p_T$ Spectra for pp Collisions at LHC}
\author{Cheuk-Yin Wong\thanks{e-mail:wongc@ornl.gov}
\address{Physics Division, Oak Ridge National Laboratory, Oak
Ridge, TN 37831} \and Grzegorz Wilk\thanks{e-mail:
wilk@fuw.edu.pl}
\address{ National Centre for Nuclear Research; Warsaw, Poland}}

\maketitle

\begin{abstract}
Phenomenological Tsallis fits to the CMS and ATLAS transverse spectra
of charged particles were found to extend for $p_T$ from 0.5 to 181
GeV in $pp$ collisions at LHC at $\sqrt{s}=$7 TeV, and for $p_T$ from
0.5 to 31 GeV at $\sqrt{s}=$0.9 TeV.  The simplicity of the Tsallis
parametrization and the large range of the fitting transverse momentum
raise questions on the physical meaning of the degrees of freedom that
enter into the Tsallis distribution or $q$-statistics.
\end{abstract}


\vspace{1cm}

Recently there is a lot of interest in the Tsallis fit to the
transverse momentum data of charged particles measured at very high
energies of RHIC and LHC experiments
\cite{STAR,PHENIX,ATLAS,CMS11,CMS12a,ALICE}. By this one understands
that the use of the Tsallis distribution with a normalization constant
$C_q$
\begin{equation}
h_q\left( p_T\right) = C_q\left[ 1 -
(1-q)\frac{p_T}{T}\right]^{\frac{1}{1-q}},\label{eq:Tsallis}
\end{equation}
describes the experimental transverse momentum distribution data.
The Tsallis distribution
 can be regarded
as a generalization of the usual exponential (Boltzmann-Gibbs)
distribution, and converges to it when the parameter $q$ tends to unity
\cite{Tsallis},
\begin{equation}
 h_q\left( p_T\right)
 \stackrel{q
\rightarrow 1}{\Longrightarrow} C_1 \exp
\left(-\frac{p_T}{T}\right).
\end{equation}
 This approach is a nonextensive generalization of
the usual statistical mechanics (characterized by a new parameter
$q$, the nonextensivity parameter) and has been very successful in
describing very different physical systems in terms of statistical
approach, including multiparticle production processes at lower
energies\footnote{For a summary of earlier attempts of using
Tsallis fits and detailed explanations of the possible meaning of
the $q$ parameter, together with up-to-date literature on this
subject, see \cite{Wil12a,Urm12,ClW}.}.

On the other hand, long time ago Hagedorn proposed the {\it QCD
inspired} empirical formula describing the data of the invariant
cross section of hadrons as a function of $p_T$ over a wide range
\cite{H}:
\begin{eqnarray}
  E\frac{d^3\sigma}{d^3p} = C \left( 1 + \frac{p_T}{p_0}\right)^{-n}
  \longrightarrow
  \left\{
 \begin{array}{l}
  \exp\left(-\frac{n p_T}{p_0}\right)\quad \, \, \, {\rm for}\ p_T \to 0, \smallskip\\
  \left(\frac{p_0}{p_T}\right)^{n}\qquad \qquad{\rm for}\ p_T \to \infty,
 \end{array}
 \right .
 \label{eq:H2}
\end{eqnarray}
where  $C$, $p_0$,  and $n$ are fitting parameters. This becomes a purely
exponential function for small $p_T$ and a purely power law function for large
$p_T$ values\footnote{Actually this QCD inspired formula was proposed
earlier in \cite{CM,UA1}.}. It coincides with Eq.
(\ref{eq:Tsallis}) for
\begin{equation}
n = \frac{1}{q - 1}\quad {\rm and}\quad p_0 = \frac{T}{q -
1}.\label{eq:coincides}
\end{equation}
Usually, both formulas are treated as equivalent from the point of
view of phenomenological fits and are often used
interchangeably \cite{STAR,PHENIX,ATLAS,CMS11,CMS12a,ALICE}. We
follow this attitude for a while and shall discuss the possible physical
implications later.

For phenomenological as well as theoretical interests, it is useful to
explore where the Tsallis fit begins to fail at higher and higher
$p_T$. Here we concentrate only on the recent high-$p_T$ data of CMS
\cite{CMS12a} and ATLAS \cite{ATLAS} Collaborations at LHC.  Excellent
fit to the $p_T$ spectra was obtained there with the Tsallis and/or
Hagedorn distributions for $p_T$ from 0.5 GeV up to 6 GeV, in $pp$
collisions at $\sqrt{s}=$7 TeV. However, CMS data extend to much
higher range of $p_T$, up to $\sim$200 GeV/c.  It will be interesting
to know whether the good Tsallis fit continues to higher transverse
momenta and how it would relate to results of fits to data at the
lower energy of $\sqrt{s}=$0.9 TeV.

\begin{figure} [h]
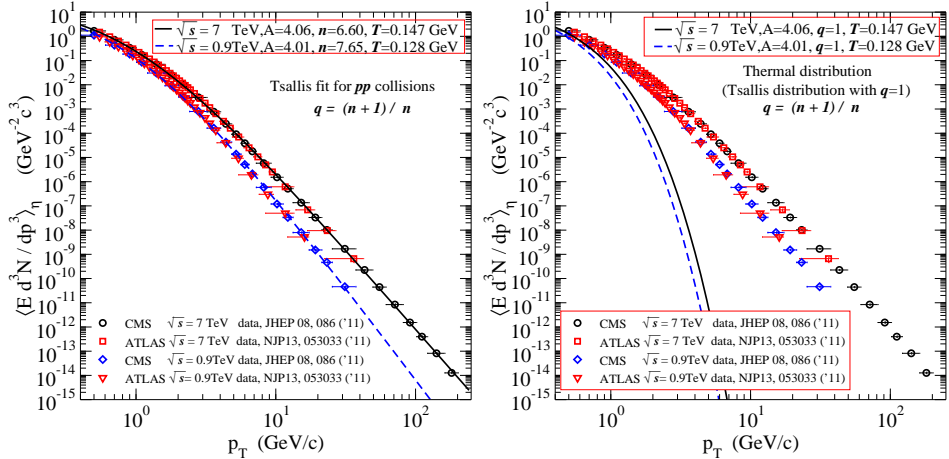

%\vspace*{-2.5cm}
\includegraphics[scale=0.34]{ptfitcmsatlas}
\includegraphics[scale=0.34]{ptfitcmsatlasq1}
\caption{ (Color online) Left panel: Tsallis fits (\ref{eq:6}) to
the CMS \cite{CMS12a} and ATLAS \cite{ATLAS} Collaborations data
for $pp$ at 7 and 0.9 TeV \cite{CMS12a}. Right panel: the same data compared
with the corresponding $q = 1$ (or $n \rightarrow \infty$)
curves.} \label{Fig1}
\end{figure}

The CMS \cite{CMS12a} and ATLAS \cite{ATLAS} $\langle Ed^3N_{\rm
  ch}/dp^3\rangle_\eta$ data shown in Fig. \ref{Fig1} are taken
essentially at the same kinematical windows, namely they correspond to
an average over the data from $\eta$=$-\eta_0$ to $\eta$=$+\eta_0$
with $\eta_0 = 2.4$ for the CMS measurements, and $2.5$ for the ATLAS
measurements\footnote{We do not include here ALICE data \cite{ALICE}
  because they are for a smaller window, $-0.8 < \eta < 0.8$, and the
  data points are slightly higher for large $p_T$ because of the
  slight $\eta$ dependence of the spectra\cite{ALICE}.}. To get the
theoretical results from the distribution for comparison with
experimental data we, therefore, calculate numerically
\begin{eqnarray}
\left \langle E\frac{d^3 N_{\rm ch}}{dp^3} \right \rangle_{\!\!\!\eta}
=\frac{1}{2\eta_{0}}\int_{-\eta_0}^{\eta_0} d\eta \frac{dy}{d\eta} \left (E\frac{d^3
N_{\rm ch}}{dp^3}\right ). \label{eq:1}
\end{eqnarray}
Here
\begin{eqnarray}
\frac{dy}{d\eta}(\eta,p_T)=\sqrt{ 1 - \frac{m^2}{m_T^2 \cosh^2
y}}~, \label{eq:2}
\end{eqnarray}
which is from Eq. (2.31) of \cite{Won94}, and $y$ is a function of
$\eta$ and $p_T$  (Eq. 2.29) of  \cite{Won94},
\begin{eqnarray}
y=\frac{1}{2} \ln \left [ \frac {\sqrt{p_T^2 \cosh^2 \eta +m^2} +
p_T \sinh \eta} {\sqrt{p_T^2 \cosh^2 \eta +m^2} - p_T \sinh
\eta}\right ]. \label{eq:3}
\end{eqnarray}
To provide a theoretical fit to the experimental data, we follow the CMS
Collaboration \cite{CMS11} and consider  the differential cross
section with a transverse  Tsallis distribution \cite{Tsallis,Wil07}
in the form
\begin{eqnarray}
E \frac{d^3N_{\rm ch}}{dp^3} =C \frac{dN_{ch}}{dy} \left (
1+\frac{E_T}{nT}\right )^{-n}, \label{eq:4}
\end{eqnarray}
where
\begin{eqnarray}
 E_T=\sqrt{m^2+p_T^2} - m, \label{eq:5}
\end{eqnarray}
and we assume $m=m_\pi=0.14$ GeV. If we assume now a rapidity plateau
structure with a constant $CdN_{\rm ch}/dy$, then the integral is
\begin{eqnarray}
\left \langle E\frac{d^3 N_{\rm ch}}{dp^3} \right \rangle_{\!\!\!\eta}
 =\frac{C}{2\eta_0}
\frac{dN_{\rm ch}}{dy}\int_{-\eta_0}^{\eta_0} d\eta \frac{dy}{d\eta}
 \left ( 1+\frac{E_T}{nT}\right )^{-n}. \label{eq:6}
\end{eqnarray}
For each value of $p_T$, the curves in Fig. \ref{Fig1} are
obtained from such a numerical  integration over $\eta$.  The
$p_T$ spectrum is therefore described by an overall constant
$A=CdN_{\rm ch}/dy$ and the parameters  $n$ and $T$.

Previously, excellent Tsallis fit to the CMS $p_T$ spectra at
$\sqrt{s}=7$ TeV was obtained for $0.5 < p_T < 6$ GeV/c by using
$n=6.6$ and $T=0.145$ GeV \cite{CMS11}. By using this set of
parameters as initial guess, we search for the fits to the spectrum
from 0.5 GeV up to 181 GeV.  We find that for the set including the
data points at the higher $p_T$ region, the best fit is obtained with
$A=$4.06, $n$=6.6, and $T=0.147$ GeV, which is essentially the same as
the set obtained previously in \cite{CMS11} (Fig. \ref{Fig1}, left
panel).  What is surprising is that the Tsallis distribution that
describes the data well at lower $p_T$ can describe the CMS data at
high $p_T$ just as well.  There is no departure of data from the
Tsallis distribution from $p_T$=0.5 GeV up to $p_T$$\sim$200 GeV.  For
additional comparison, we include the ATLAS data in
Fig. \ref{Fig1}. The ATLAS measurement has a slightly larger $\eta$
window, $|\eta|_{\rm ATLAS}\le 2.5$, instead of CMS's $|\eta|_{\rm
  CMS}\le 2.4$, and its $p_T$ values extends from 0.5 GeV/c to 36
GeV/c for $\sqrt{s}$=7 TeV.  The ATLAS data at $\sqrt{s}$=7 TeV are
consistent with the CMS data and the Tsallis fit (Fig. \ref{Fig1},
left panel).

On the other hand, at the lower energy of $\sqrt{s}=0.9$ TeV, we
find that the $pp$ data can be described well by the higher value
of $n$=7.65 (with $A$=4.01 and $T$=0.128 GeV) in Fig.\ref{Fig1}.
Again, the CMS data and the ATLAS data are consistent with each
other. The Tsallis distribution  gives a good description of the
$\sqrt{s}$=0.9 TeV data from $p_T$=0.5 GeV/c  to  $p_T$=31 GeV/c
(Fig. \ref{Fig1}, left panel) .

It is instructive to visualize the difference between the Tsallis
distribution and the corresponding thermal distribution
characterized by the same temperature $T$ and by $q = 1$ (or $n\to
\infty$).  This is shown on the right panel of Fig. \ref{Fig1}.
Changing only the temperature parameter $T$ would not give a
better fit to the experimental data, to do this it is necessary to
allow $q$ to vary and become $q > 1$.

The overall good agreement of the Tsallis parametrization and the
experimental data is amazingly good. The absence of a departure of
data from the Tsallis distribution from $p_T$=0.5 GeV to 181 GeV
indicates that there are essentially only three degrees of freedom
that count for the description of the $p_T$ distribution: an
overall magnitude, and two other degrees of freedom to describe
the shape. This result should be confronted with the (apparently
equally successful) many parameter fits also presented in
\cite{ATLAS,CMS11,CMS12a,ALICE} and using known Monte Carlo
programs. It can be interpreted as indication that, whereas
hadronizing system formed in the process of particle production is
very complex, only few degrees of freedom are really important. At
lower energies (and for lower values of $p_T$) this was regarded
as indication that such system can be described by simple (with
$q=1$) or generalized (with $q > 1$) statistical models
\cite{Wil07,Wil12a}. The results presented here indicate that,
either such models work also for such large $p_T$ or there are
some specific, dynamical rather than purely statistical, phenomena
at work as advocated recently in \cite{Wil12b}. In any case, we
need to take the physics contents of the Tsallis model seriously.

At this point let us come back to formula (\ref{eq:H2}) (or
(\ref{eq:6})). It was proposed in \cite{CM} long time before
Tsallis works \cite{Tsallis} (and followed later by \cite{UA1} and
\cite{H}) with a simple aim: to phenomenologically interpolate
between the {\it soft region} of $p_T$$ \rightarrow$0,
characterized by exponential behavior of $p_T$ distributions, and
the {\it hard region} of $p_T$$\gg$1 GeV, believed to be properly
described by QCD. However at that time, the exponent index
obtained from fits to $p_T$ data was equal to $n$$\sim$8 (or
bigger), far away from the expected point interaction value of
$n$$\sim$4 \cite{Bla75,Bla77,CountR}. As one can see from our
fits, this region of dominance of truly point-like hard
interactions is still far away (albeit $n$ diminishes noticeably
between $900$ GeV and $7$ TeV). Actually, comparing our results
with compilation of results at lower energies provided by
\cite{Wibig} one observes that the naive counting rule result of
$n=4-6$ \cite{CountR} seems to be out of reach. It means
therefore, that even at the highest energies, we do not deal with
the point-like objects expected from the naive field theory
expectations and there is always an additional (to that resulting
from a hard collision) $p_T$ transfer, perhaps preceded by a kind
of the multiple scattering process or constituent scattering which
make the finally observed spectra softer than naively expected.

It is of interest to note in this connection that the power of
$n\sim 6.6 -  7.6$ in the transverse momentum spectrum at these
high energies may be related to the constituent interchange model
of Blankenbecler and Brodsky and Gunion \cite{Bla75}.  In the
basic quark models diagrams, the power index in $p_T$ dependence
of the inclusive spectra can be inferred from the counting rule
involving the collision of the active constituents
\cite{Bla75,Bla77} (for a review, see \cite{Won94}).   If one
assumes that the dominant basic  high-$p_T$ process in $pp \to \pi
X$ comes from  $qq \to qq$, then the counting rule gives a
transverse momentum dependence of $1/(p_T^2)^2$ with $n=4$, which
differ from the observed power index, as we mentioned earlier.    On
the other hand, if one assumes that the basic process is
$q+$meson$\to$$q$+meson, then the counting rule gives $n=8$ which
is close to the observed value. The basic process of
$q$+meson$\to$$q$+meson  may appear  dominant because of the
strong quark-hadron coupling.

We close with the following observation. From what was shown here
it seems that phenomenologically the two-parameter, QCD-inspired
formula (\ref{eq:4}) is as good as the two-parameter Tsallis
formula with $n \rightarrow 1/(q - 1)$. The only difference is in
the interpretation, i.e., in the possible thermodynamical origin
(among others) of the Tsallis formula \cite{Wil12a,Urm12,ClW}. In
this case Tsallis fits would cover the whole energy range of
experiments uniformly interpreted in terms of thermal (extensive
or nonextensive) model. That this view is reasonable was shown in
papers explicitly demonstrating that nonextensive-thermodynamics
satisfies all demands of the usual thermodynamics applied to
systems that posses intrinsic fluctuations, memory effects, are
limited and/or nonhomogeneous etc. \cite{View1,View2,View3}.
Whether the recent CMS result presented here fits to this picture
or rather calls for some novel explanation of Tsallis formula (and
parameter $q$) remains, however, for a time being an open
question.

\vspace*{0.3cm} \centerline{\bf Acknowledgment}

\vspace*{0.3cm}   The research  was supported in part by the
Division of Nuclear Physics, U.S. Department of Energy (CYW) and
by the Ministry of Science and Higher Education under contract
DPN/N97/CERN/2009 (GW).

\end{document}